\documentclass[runningheads]{llncs}

\usepackage[T1]{fontenc}
\usepackage{graphicx,booktabs}
\usepackage{amssymb,mathtools,amsmath}
\usepackage{mathbbol}
\usepackage{listings}
\usepackage{lstcustom}
\usepackage{tikz}
\usetikzlibrary{automata, positioning, arrows}
\usepackage{macros}
\usepackage{comment}
\usepackage{todonotes}
\usepackage{jmllistings}
\usepackage[inline]{enumitem}
\usepackage{etoolbox}

\newtoggle{extended}
\toggletrue{extended} 
\lstdefinelanguage{stipula}{%
  basicstyle=\ttfamily,keywords={stipula,assets,fields,agreement,now,if,else,false,true}
}

\def\stipula#1{\mbox{\lstinline[mathescape=true,language=stipula]!#1!}}
\def\jml#1{\mbox{\lstinline[mathescape=true,language={[JML]Java}]|#1|}}

\AtBeginDocument{}

\renewcommand{\todo}[1]{}

\begin{document}

\title{Formal Verification of Legal Contracts:\\
  A Translation-based Approach}

\titlerunning{Formal Verification of Legal Contracts}

\author{Reiner H\"ahnle\inst{1}\orcidID{0000-0001-8000-7613} \and
  Cosimo Laneve\inst{2}\orcidID{0000-0002-0052-4061} \and
  Adele Veschetti\inst{1}\orcidID{0000-0002-0403-1889}}

\authorrunning{R.~H\"ahnle, C. Laneve, A. Veschetti}

\institute{
  Department of Computer Science, TU Darmstadt, Germany\\
  \email{\{reiner.haehnle,adele.veschetti\}@tu-darmstadt.de}
  \and
  Department of Computer Science, University of Bologna, Italy\\
  \email{cosimo.laneve@unibo.it}
  }

\maketitle              

\begin{abstract}%
  {\Stipula} is a domain-specific programming language designed to model
  legal contracts with enforceable properties, especially those
  involving asset transfers and obligations.
  This paper presents a
  methodology to formally verify the correctness of Stipula contracts
  through translation into Java code annotated with Java Modeling
  Language specifications. As a verification backend, the deductive
  verification tool KeY is used. Both, the translation and the
  verification of partial and total correctness for a large subset of {\Stipula} contracts, those with disjoint cycles, is fully automatic.
  %
Our work demonstrates that a general-purpose
  deductive verification tool can be used successfully in a
  translation approach.

\end{abstract}

\section{Introduction}
\label{sec:introduction}

As the legal domain continues its digital transformation, the demand
for precise, machine-verifiable representations of legal contracts
becomes increasingly important. Traditional legal texts, written in
natural language, are inherently ambiguous and prone to
misinterpretation, making them challenging to process, analyze, or
verify by automated means. To address this issue, several projects are
being developed for defining programming languages to write legal
contracts, \emph{e.g.}~\cite{Lexon,Catala,Accord,OpenLaw}.  Although
these projects introduce a precise syntax for legal contracts and
provide graphical tools to associate normative elements with code,
they pay limited attention to the verification of correctness. Yet,
this aspect is crucial in legal contexts, where ambiguities or
unintended behavior during execution can lead to significant legal and
financial repercussions. Despite advances in the formal verification
of legal contracts, formal reasoning is rare in practice due to tool
complexity.

To overcome this lack of automatic verification techniques, in 2021,
we designed {\Stipula}~\cite{CrafaL20,CrafaLSV23,HOStipula}, a new
domain-specific language with few concise and intelligible primitives
that have a precise correspondence with the distinctive elements of
legal contracts. The language has a formal operational semantics, so
that the behavior is fully specified and amenable to automatic
verification. Its current tool chain~\cite{stipulaprototype} contains a
runtime environment, a type checker, a graphical IDE, and an analyzer
verifying the reachability of clauses~\cite{TecReport2025,Laneve2024}.

In this paper, we advance the automatic verification of {\Stipula}
contracts by introducing a systematic approach for verifying contracts
through translation into a general-purpose deductive verification
tool.
%
%
%
%
%
%
%
Specifically, we translate {\Stipula} contracts into Java
programs annotated with JML (Java Modeling Language) specifications,
and we employ the {\KeY} verification system
\cite{DBLP:series/lncs/10001,KeYTutorial24} as a backend.
We define translation patterns and principles that preserve the
semantics of core {\Stipula} constructs --- such as asset transfers,
functions, and events --- within the program logic supported by
{\KeY}. Our approach admits both manual and automated translation, and
it accommodates a variety of contract types, including those with
cyclic behavior and partial asset transfers.

It is by no means evident that a translation-based approach to
verification will succeed. For deductive verification tools such as
{\KeY} to operate in a fully automated mode, the availability of
sufficiently precise specification annotations is crucial.  In the
general case, such specifications need to be provided manually, which
renders deductive verification inherently interactive and,
consequently, costly.  \todo{Reviewer 2, comment 4} In the case of
{\Stipula}, every smart contract is associated with a well-defined
automaton: the states represent control points of the contract, while
the transitions correspond to its clauses, such as functions or
events. This automaton-based representation makes the contract's
behavior explicit and provides a natural foundation for reasoning
about possible execution paths.  Moreover, when the automaton of a
contract satisfies the structural property of \emph{disjoint cycles}
-- that is, cycles that do not share any states -- it becomes possible
to automatically synthesize suitable JML annotations, including
guarantees and loop invariants.  This synthesis step significantly
reduces the annotation burden and paves the way for a high degree of
automation in the verification process, thus demonstrating the
potential of the translation-based approach in practice.

Nonetheless, there are limitations that affect full generality.
Time-dependent behavior must be scheduled symbolically and evaluated
statically.  Our technique handles this by using symbolic boolean
variables representing the time constraint. The event is then
translated into a conditional guarded by the symbolic variable that
{\KeY} evaluates by exploring both possibilities: one where the event
occurs and one where it does not.  More complex forms of time
management, such as dynamically registered events during loops or
asynchronous behavior, are outside the current scope.  Similarly,
non-deterministic behavior, such as contracts where multiple
transitions may be enabled concurrently, is not yet supported: our
translation assumes that guards deterministically select a unique
transition.  These aspects pose challenges for extending the approach
beyond the verified fragment we currently target.

The paper is structured as follows.  Section~\ref{sect:background}
introduces the {\Stipula} language and its execution semantics; an
illustrative example highlights how obligations, permissions, and
events are expressed in {\Stipula}.  We also provide a short
introduction to {\KeY}.
Section~\ref{sec:translation} formalizes our translation methodology
from {\Stipula} to Java, detailing how fields, assets, functions,
events, and contract behavior are encoded in Java and specified with
JML.
Section~\ref{sect:casestudies} discusses the implementation of our
translator tool and presents case studies that apply our approach to
representative {\Stipula} contracts, demonstrating the effectiveness
of the method and discussing verification outcomes.
Section~\ref{sect:relatedworks} reviews related efforts in contract
languages and formal verification.  Section~\ref{sect:conclusion}
concludes the paper and outlines directions for future work.

\section{Background}
\label{sect:background}

To set the stage for our translation-based verification method, we
briefly review the relevant background. Section \ref{subsect:stipula}
presents the {\Stipula} language, emphasizing its constructs for
assets, states, and timed events.  Section \ref{subsect:key}
introduces the {\KeY} system, which we later employ as the
verification backend.

\subsection{{\Stipula}}
\label{subsect:stipula}

%
%
%
A {\Stipula} contract consists of a set of parties, states, assets,
fields, and a set of functions and events, generically called
\emph{clauses}. The declaration of a contract is defined in
Fig.~\ref{tab:stipula-syntax}, where \stipula{C} is the name of the
contract, $\vect{\tt h}$ and $\vect{\tt x}$ are the \emph{assets} and
\emph{fields}, respectively, $\vect{\tt A}$ are the \emph{parties}.
\begin{figure}[h]
{\small
\[
\begin{array}{l}
\stipula{stipula C \{}
\\
\qquad \stipula{asset} \; \vect{\tt h}
\\
\qquad \stipula{field} \; \vect{\tt x} 
\\
\qquad \stipula{agreement(} \vect{\tt A} \stipula{) \{  }
\\
\qquad \qquad \vect{{\tt A}_1} \; : \; \vect{{\tt x}_1}
\\
\qquad \qquad \cdots \quad {\tt //} \; \bigcup_{i \in 1..n} \vect{{\tt A}_i} \subseteq \vect{{\tt A}}   \; , \quad  \bigcup_{i \in 1..n} \vect{{\tt x}_i} \subseteq\vect{{\tt x}} \; , \quad	\bigcap_{i \in 1..n} \vect{{\tt x}_i} = \varnothing
\\
\qquad \qquad  \vect{{\tt A}_n} \; : \; \vect{{\tt x}_n}
\\
\qquad \stipula{\}} \tostate \Qwithat  
\\
\qquad     F
\\
\stipula{\}}
\end{array}
\]
%
%
\\ 
\[
\begin{array}{l@{\quad}l@{\quad}l}
\mathit{Functions} & F \; ::= & \zero  \quad | \quad   \Qwithat \;\A : \; {\sf f}(\vect{\sf y})[\vect{\sf k}]\,(E)\{
\,S\, ; \, W\,\} \,\tostate\, \Qwithat'\;  F
\\
\mathit{Prefixes} & P \; ::= & E \, \send \, {\sf x} \quad | \quad E \send \, \A \quad | \quad E \lolli {\sf h}, {\sf h}' 
\quad |\quad E \lolli {\sf h} ,  \A \\
\mathit{Statements} & S \; ::= &  \zero  \quad | \quad P \; S \quad | \quad \ifte{E}{S}{S} \; S
\\
\mathit{Events} &  W \; ::= & \zero  \quad | \quad \stipula{now}+k \,\event\, \Qwithat\, \{ \, S \,\} \,\tostate\, \Qwithat'\; \;  W 
\\
\mathit{Expressions }& E \; ::= &  v \quad | \quad X  
\quad | \quad   E \, {\sf op} \, E \quad | \quad  {\sf uop} \, E
\\
\mathit{Values} & v ::= &  n \quad |\quad \stipula{false} \quad |\quad \stipula{true} \quad | \quad  s 
\end{array}
\]
}
\caption{\label{tab:stipula-syntax} Syntax of {\Stipula}}
\end{figure}
The \stipula{agreement} construct declares the parties that set the
initial value of the fields and the initial state of the contract. For
example, if the agreement has three parties $\A_1, \A_2, \A_3$ and the
contract has two fields $\x_1, \x_2$, if it declares $\A_1: \x_1$ and
$\A_2, \A_3 : \x_2$ then $\x_1$ will be set by $\A_1$ and $\x_2$ will be set upon
agreement on the value between $\A_2$ and $\A_3$.
When the agreement is concluded, the parties may invoke a function in
$F$.

A \emph{function}
$\Qwithat \;\A : \; {\sf f}(\vect{\sf y})[\vect{\sf k}]\,(E)\{ \,S;\
W\,\} \,\tostate\, \Qwithat'$ can be invoked by a party $\A$ if the
contract is in state $\Q$ and the guard $E$ is \emph{true}.  The names
$\vect{\sf y}$ and $\vect{\sf k}$ are the formal parameters of {\tt
  f}; they are kept separate because $\vect{\sf y}$ are field values
while $\vect{\sf k}$ are asset quantities.

Function bodies are \emph{statements} followed by \emph{events}.  The
former include value transfers, asset movements, conditional logic,
and field assignments. {\Stipula} distinguishes between different
transfer operations: field and message updates use the symbol
$\rightarrow$ with the usual semantics of assignment, while asset
transfers use the linear implication operator $\lolli$ to emphasize
the conservation semantics. For instance, an expression like
``$1 \lolli \stipula{wallet}, \stipula{Seller}$'' denotes exclusive
transfer of a unit in {\tt wallet} to the {\tt Seller} and, \emph{at
  the same time}, the {\tt wallet} is decreased by 1.  In contrast
``\stipula{code} $\rightarrow$ \stipula{Licensee}'' models
non-exclusive passing of information. The operation
``$\stipula{wallet} \lolli \stipula{wallet}, \stipula{Seller}$'' is
always shortened to ``$\stipula{wallet} \lolli \stipula{Seller}$''.

Events
$\stipula{now}+k \,\event\, \Qwithat\, \{ \, S \,\} \,\tostate\,
\Qwithat'$, where $k$ is either a natural number or a field name,
define a statement $S$ to be executed if, \emph{after} $k$ time units
from the current execution, the contract is in the state $\Q$. If the
event is executed, the contract will transition to the state $\Q'$.

We refer to article~\cite{CrafaLSV23} for background on the design of
{\Stipula}, as well as its formal semantics.  A comment about the
model of time in {\Stipula} may be useful for what follows.  The model
has a multiset of events to be executed; every event has a time value
that is a natural number representing minutes. This number is computed
when the event is created by replacing \stipula{now} with 0 in the
expression $\stipula{now}+k$ (recall that $k$ is either a natural
number or a field name).
Time advances when the contract has no statements to execute and no
events can be triggered. In such cases, a ``tick'' occurs,
decrementing the time values in the multiset of events. Events with
negative time values are discarded.  Subsequently, any event whose
initial state matches the current contract state and whose time value
reaches zero may be scheduled for execution.%
\footnote{The syntax of {\Stipula} in~\cite{CrafaLSV23} admits
  absolute time expressions like {\tt "2022/1/1:00:15"}; these
  expressions are rewritten into terms $\stipula{now}+k$ when the
  event is created.}

In this paper we use the state transition models of {\Stipula}
contracts, called the \emph{underlying automata}. The states of these
automata are those of the {\Stipula} contract. The transitions
correspond to clauses and are labelled either with the function name
(the party name is always omitted, for simplicity's sake we assume
that function names are pairwise different) or with the event line
number (\emph{e.g.}, \stipula{ev}$_{10}$ is the event at code line
10). Fig.~\ref{fig:license-fsm} shows an underlying automaton.

We illustrate {\Stipula} through two representative examples that
allow us to highlight crucial {\Stipula} features:
\begin{description}
\item[\emph{normative permissions}:] functions are enabled only for
  specific parties in specific states;
\item[\emph{asset safety}:] assets are never duplicated or lost;
  transfers are explicitly encoded and conditional;
\item[\emph{timed obligations}:] the \stipula{event} construct encodes
  deadlines and enforces compliance without external intervention;
\item[\emph{stateful logic}:] contract progress is encoded through
  explicit state transitions, supporting both branching and linear
  workflows.
\end{description}

%
\begin{example}[The License Contract]
  Listing~\ref{code:license-stipula} defines a \emph{license contract}
  that regulates a licensing transaction between a \stipula{Licensor}
  and a \stipula{Licensee}, with time-bound trial periods and the
  possibility to purchase or decline the license.
  In particular, the \stipula{Licensee} may request a trial and then
  decide whether to buy the license.  If the \stipula{Licensee} does
  not purchase the license before the trial period expires, the
  contract terminates and the cost is automatically returned to the
  \stipula{Licensee}.

  The contract begins with an \stipula{agreement} clause in
  Line~\ref{line:lic:agb}, where both parties define the trial start
  time (\stipula{t\_start}), its duration (\stipula{t\_limit}), and
  the license \stipula{cost}.  This mirrors the legal principle of
  mutual consent (``meeting of the minds''): no contract behavior is
  enabled until consensus is reached. Upon agreement, the first
  control state is \stipula{@Init} (Line~\ref{line:lic:age}).

{\small
\begin{lstlisting}[numbers=left,frame=none,xleftmargin=2.5em,mathescape=true,language=stipula,
caption={The License contract in {\Stipula}},label={code:license-stipula},escapechar=§]
stipula License {
    asset balance, token
    field t_start, t_limit, cost, code
    agreement (Licensor, Licensee)(t_start, t_limit, cost) { §\label{line:lic:agb}§
        Licensor, Licensee : t_start, t_limit, cost
    } $\tostate$ @Init §\label{line:lic:age}§
    @Init Licensor: offer(x)[n] {
        n $\lolli$ token
        x $\send$ code
        now + t_start $\event$ @Prop { token $\lolli$ Licensor } $\tostate$ @End
    } $\tostate$ @Prop
    @Prop Licensee: activate()[b] (b == cost) {
        b $\lolli$ balance
        code $\send$ Licensee ;
        now + t_limit $\event$ @Trial {
                balance $\lolli$ Licensee
                token $\lolli$ Licensor
                } $\tostate$ @End
    } $\tostate$ @Trial
    @Trial Licensee: buy()[] {
        balance $\lolli$ Licensor
        token $\lolli$ Licensee
    } $\tostate$ @End
}
\end{lstlisting}}

The \emph{underlying automaton} of \stipula{License} is shown in
Fig.~\ref{fig:license-fsm}.  
\begin{figure}
\centering
\scalebox{0.75}{
\begin{tikzpicture}[->, >=stealth', node distance=2.5cm, every state/.style={draw, minimum size=1.2cm},initial text=]
  \node[state, initial] (init) {\stipula{Init}};
  \node[state, right of=init] (first) {\stipula{Prop}};
  \node[state, right of=first] (run) {\stipula{Trial}};
  \node[state, below of=first] (end) {\stipula{End}};

  \path (init) edge node[above] {\stipula{offer}} (first)
        (first) edge node[above] {\stipula{activate}} (run)
        (first) edge node[left] {\stipula{ev}$_{1}$} (end)
        (run) edge node[left] {\stipula{buy}} (end)
        (run.south) edge node[right] {\stipula{ev}$_{2}$} (end.east);
\end{tikzpicture}
}
\caption{Underlying automaton of the \stipula{License} contract}
\label{fig:license-fsm}
\end{figure}
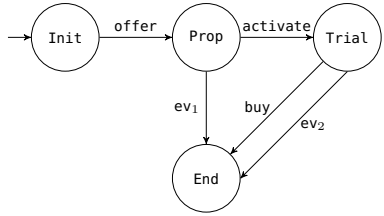
In the \stipula{Init} state, the \stipula{Licensor} may invoke
\stipula{offer}, transferring a \stipula{token} (representing the
license) into escrow and generating a license \stipula{code}.  A
scheduled \stipula{event} is simultaneously registered at
\stipula{t_start} time units in the future to reclaim the token if the
\stipula{Licensee} fails to act within the trial start window.  The
contract then moves to \stipula{Prop}.
If no function is called in this state, then time is advanced by one
unit, eventually triggering the timeout at \stipula{t_start}.
These event transitions are dynamically added to the state machine
during function execution, \emph{e.g.}, the two events in
Fig.~\ref{fig:license-fsm} are not initially present.  This dynamic
creation increases the expressiveness of
{\Stipula}~\cite{TecReport2025}.

In \stipula{Prop}, the \stipula{Licensee} can activate the trial by
paying the \stipula{cost} (observe that \stipula{b} is an asset of the
\stipula{Licensee}), which is transferred to the contract's
\stipula{balance}: the contract acts as a notary for assets that are
not finally disposed.  Now the license code is revealed to the
\stipula{Licensee} and another \stipula{event} is scheduled to handle
the expiration of the trial period: if the license is not purchased
before \stipula{t_limit} time units, the balance is refunded and the
token is returned to the \stipula{Licensor}.

  In the \stipula{Trial} state, the \stipula{Licensee} has the option
  to purchase the license via \stipula{buy}, which finalizes the
  transaction by transferring the \stipula{balance} to the
  \stipula{Licensor} and assigning the \stipula{token} permanently to
  the \stipula{Licensee}.  In either case, the contract terminates in
  the \stipula{End} state, without retaining any asset.
\end{example}

\begin{example}[The Deposit Contract]
  \label{ex:DepositContract}
  Listing~\ref{code:deposit-stipula} presents a \emph{deposit
    contract} that models the interaction between a \stipula{Farm} and
  a \stipula{Client}.  The \stipula{Farm} deposits flour, while the
  \stipula{Client} purchases and withdraws the corresponding amount at
  an agreed price.  The contractual terms are enforced over a validity
  period of 365 days.  The clause \stipula{send} allows the
  \stipula{Farm} to deposit flour into the contract's stock.  In
  \stipula{send()[h]}, the assignment \stipula{h $\send$ Client} is
  only an informational message to the \stipula{Client} about the
  deposited amount (\stipula{h} is not emptied), while the asset
  transfer \stipula{h $\lolli$ flour} increases the contract's
  internal balance of flour (and, at the same time, empties the asset
  \stipula{h}). Observe that the converse ordering of the instructions
  (\stipula{h $\lolli$ flour $\quad$ h $\send$ Client}) sends 0 to the
  \stipula{Client} as informational message.  Also, no flour is
  transferred to the \stipula{Client} by send.  Actual delivery occurs
  only with \stipula{buy()[w]}, where \stipula{(w/cost_flour) $\lolli$
    flour, Client} transfers flour from the contract's stock to the
  \stipula{Client} (hence \stipula{w/cost_flour} must not be greater
  than \stipula{flour}) and \stipula{w $\lolli$ Farm} represents the
  payment to the \stipula{Farm}.  In this way, send supplies the
  stock, whereas buy withdraws from it under payment.

    {\small
\begin{lstlisting}[numbers=left,frame=none,xleftmargin=2.5em,mathescape=true,language=stipula,
caption={The \stipula{Deposit} contract in {\Stipula}},label={code:deposit-stipula},escapechar=§]
stipula Deposit {
    asset flour
    field cost_flour
    agreement (Client, Farm)(cost_flour) {
        Client, Farm : cost_flour
    } $\tostate$ @Start
    @Start Farm : begin()[h]{
        h $\send$ Client
        h $\lolli$ flour;
        now + 365 $\event$ @RunF { flour $\lolli$ Farm } $\tostate$ @End
        now + 365 $\event$ @RunC { flour $\lolli$ Farm } $\tostate$ @End
    } $\tostate$ @RunC
    @RunF Farm : send()[h]{
        h $\send$ Client
        h $\lolli$ flour
    } $\tostate$ @RunC
    @RunC Client : buy()[w](w/cost_flour <= flour){
        (w/cost_flour) $\lolli$ flour, Client
        w $\lolli$ Farm
    } $\tostate$ @RunF
}
\end{lstlisting}}
\begin{figure}[h]
\centering
\scalebox{0.75}{
\begin{tikzpicture}[->, >=stealth', node distance=2.5cm, every state/.style={draw, minimum size=1.2cm}, initial text=]
  \node[state, initial] (start) {\stipula{Start}};
  \node[state, right of=start] (runc) {\stipula{RunC}};
  \node[state, right of=start, yshift=-2.5cm] (runf) {\stipula{RunF}};
  \node[state, right of=runc] (end) {\stipula{End}};

  \path (start) edge node[above, sloped] {\stipula{begin}} (runc)
        (runc) edge[bend left=30] node[left] {\stipula{buy}} (runf)
        (runf) edge[bend left=30] node[left] {\stipula{send}} (runc)
        (runc) edge node[above, sloped] {\stipula{ev}$_{1}$} (end)
        (runf) edge node[below, sloped] {\stipula{ev}$_{2}$} (end);
\end{tikzpicture}
}
\caption{Underlying automaton of the \stipula{Deposit} contract}
\label{fig:deposit-fsm}
\end{figure}
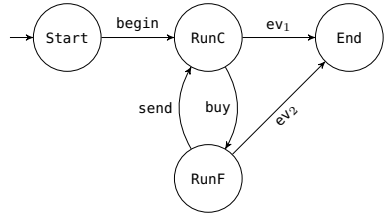

Unlike the \stipula{License} contract, the \stipula{Deposit} contract
has cyclic interactions: the \stipula{Farm} and \stipula{Client} can
repeatedly invoke \stipula{send} and \stipula{buy} as long as the
contract remains valid, as shown in the automaton in
Fig.~\ref{fig:deposit-fsm}.  This is modeled by the function
\stipula{send} transitioning from \stipula{RunF} to \stipula{RunC} and
the function \stipula{buy} in the reverse direction, enabling
iteration until the validity period expires.
\end{example}

The alternation between \stipula{RunF} and \stipula{RunC} in
Example~\ref{ex:DepositContract} is by design: it ensures that each
iteration of the cycle is well-formed and contributes to disjointness
of cycles in the underlying automaton.  Allowing the \stipula{Client}
to buy repeatedly without intervening deposits would introduce
overlapping cycles, which are outside the scope of the current
automatic verification approach.  Cycles increase the complexity of
formal analysis~\cite{Laneve2024}. In particular, this feature,
combined with deadlines (the events) and permission-dependent
transitions enlarge the state space and introduce subtle cross
dependencies, making automated verification challenging.

\begin{definition}[Cyclic Contracts and Disjoint Cycles]
  \label{def:determinate}
  A {\Stipula} contract is \emph{cyclic} if its underlying automaton
  has a sequence of pairwise different states $\Q_1,\ldots,\Q_n$ with
  transitions $\Q_i \lred{\mu_i} \Q_{i+1}$ and
  $\Q_n \lred{\mu_n} \Q_{1}$ where $\mu_i$ are either function names
  or events.  The contract has \emph{disjoint cycles} if different
  cycles have no state in common.
  %
  %
\end{definition}

We focus on contracts whose automata contain only disjoint cycles ---
a structural property that simplifies path generation and enables
modular, fully automatic verification.  For example, the
\stipula{Deposit} contract includes a single cycle, while
\stipula{License} is entirely acyclic.  Disjoint cycles ensure
determinism: If a state has multiple outgoing transitions, at most one
of them belongs to any given cycle.  This allows our tool to
synthesize all execution paths statically, without ambiguity or
runtime checks.

While this restriction enhances tractability, it limits generality.
Contracts with overlapping cycles are currently out of scope, though
we discuss possible extensions in the conclusion.  The current
implementation does not yet enforce that cycles are disjoint; an
algorithm for checking it is presented in
\iftoggle{extended}{Appendix~\ref{appendix:cycles} }{the extended
  version of the paper~\cite{extended-paper}} and will be integrated
in a future version.

\subsection{Deductive Verification with the {\KeY} System}
\label{subsect:key}

The {\KeY} system \cite{DBLP:series/lncs/10001,KeYTutorial24} is a
deductive verification framework for Java that combines symbolic
execution, invariant reasoning, and method contracts\footnote{Not to
  be confused with {\Stipula} contracts. In this paper, both kinds of
  contract are featured, but it should be obvious from the context
  when we mean {\Stipula} contracts or Java/JML method contracts.} on
top of a calculus for an expressive program logic.
The main use case of {\KeY} is formal verification of Java programs
annotated with specifications written in JML. 
It provides an interactive user environment, where one can
construct correctness proofs of Java methods against their JML
contracts. These contracts typically include preconditions
(\jml{requires}), postconditions (\jml{ensures}), frame
conditions (\jml{assignable}), class invariants, and auxiliary
annotations such as loop invariants and assertions. Given a
JML-annotated Java class, {\KeY} translates the specifications into
logical proof obligations and attempts to discharge them using its
symbolic execution engine.
The details of the verification process are irrelevant for the purpose
of this paper: {\KeY} is used as a \emph{black box}, the input is a
JML-annotated Java file that results from translation of a {\Stipula}
contract. In general, {\KeY} is used in interactive or auto-active
mode, however, for all case studies discussed below, the verification
is \emph{fully automatic}.
%
%
%



\section{Translation Approach}
\label{sec:translation}

To enable formal verification of {\Stipula} contracts, we define a
systematic translation into Java code annotated with JML
specifications. Java and JML are versatile enough to express
{\Stipula} in a natural manner: functions are modeled by stateful Java
methods, the specific semantics of assets are modeled by suitable JML
method contracts, etc.; the details are given below.  Our translation
preserves the semantics of {\Stipula}'s normative constructs,
including asset ownership, state transitions, permissions, and events,
by representing them as verifiable Java/JML proof obligations.

\subsection{Assets, Fields and States}
\label{sec:no-time}

We distinguish three categories of contract data: \emph{fields},
\emph{assets}, and the \emph{control state}.  Each is translated into
corresponding elements in the target Java class, along with
accompanying JML specifications to capture correctness properties.

\paragraph{Fields.}

{\Stipula} contracts define mutable fields that store
contract-specific state information, such as numeric counters, time
stamps, or configuration parameters.  These fields evolve during
execution and influence both transitions and obligations.  We must
ensure that field updates are explicitly modeled and tracked in the
translated Java code.

\begin{definition}[Field Mapping]
  Let $\mathcal{F}$ be the set of mutable fields declared in a
  {\Stipula} contract. Each field $\stipula{f} \in \mathcal{F}$ is
  translated to a \jml{static} Java field, which holds the contract
  state data. Changes to these fields are tracked and constrained
  using JML \jml{assignable} clauses.
\end{definition}

This representation allows JML to reason about field updates and
supports the specification of frame conditions through
\jml{assignable} clauses.

\paragraph{Assets.}
In {\Stipula}, assets are classified automatically as \emph{divisible}
or \emph{indivisible} to enable explicit ownership modeling and
prevent asset duplication or loss.  The translator infers this
heuristically from usage: assets transferred with a variable quantity
(``\stipula{v $\lolli\;$ h, h'}'') are treated as divisible and
represented as an \jml{int}, while those transferred directly
(``\stipula{h $\lolli\;$ h'}'') are considered indivisible and
represented as a \jml{boolean}.  To enforce exclusivity for
indivisible assets, safety preconditions like ``\jml{requires
  !h.token}'' are generated.  Based on this classification, we now
formally define how asset ownership is represented.

\begin{definition}[Asset Ownership Representation]
  Let $\mathcal{A}$ be the set of declared assets, and $\mathcal{P}$
  the set of participants in a {\Stipula} contract \stipula{C}. Each
  asset $\stipula{a} \in \mathcal{A}$ is represented by a set of Java
  fields:
  \[
    \{ \stipula{P.a} \mid \stipula{P} \in \mathcal{P} \} \cup \{
    \stipula{C.a} \},
  \]
  where each field \stipula{P.a} is either of a numeric type
  (\texttt{int}) indicating the quantity of asset \stipula{a} held by
  party \stipula{P} or a boolean variable indicating that the party
  \stipula{P} \emph{owns} asset~\stipula{a}.  The field
  \stipula{C.a} tracks the quantity owned by the contract
  itself. 
\end{definition}

To ensure that each indivisible asset is held by exactly one owner at
all times, we define an invariant that formalizes the exclusivity
constraint.

\begin{property}[Indivisible Asset Exclusivity]
  \label{prop:asset-exclusivity}
  Each asset is owned by exactly one party in $\mathcal{P}$ or by the
  contract \stipula{C} at any given time. Formally, the following
  invariant must hold for all $\stipula{a} \in \mathcal{A}$:\quad
  $ \bigvee_{X \in \mathcal{P} \cup \{ \stipula{C} \}} \left(
    X.\stipula{a} \land \bigwedge_{Y \neq X} \lnot Y.\stipula{a}
  \right) $.

  This invariant becomes part of the translated Java class as a static
  JML \stipula{invariant}, ensuring asset linearity throughout
  execution.
\end{property}
In our current encoding, exclusivity of indivisible assets is often
enforced via explicit postconditions on methods.  Alternatively, these
properties can be stated as JML class invariants, which would provide
a more uniform and modular treatment; we plan to explore this
refinement in future versions of the tool.

\begin{example}
  The asset \stipula{token} in the License contract in
  Section~\ref{sect:background} is modeled by three Java fields:
  \jml{License.token}, \jml{Licensee.token}, and \jml{Licensor.token}.
  Property~\ref{prop:asset-exclusivity} is ensured in JML as: {\small

  \begin{center}
  \begin{tabular}{ll}
    \jml{static invariant} & \jml{(License.token$\land\lnot$Licensee.token$\land\lnot$Licensor.token)}
    \\
    & \jml{$\lor$ (Licensee.token$\land\lnot$License.token$\land\lnot$Licensor.token)}
    \\
    & \jml{$\lor$ (Licensor.token$\land\lnot$License.token$\land\lnot$Licensee.token)}
  \end{tabular}
  \end{center}
    }
\end{example}

For divisible assets, rather than enforcing exclusivity, we ensure
that the total quantity of the asset remains constant throughout the
contract execution.  This principle is captured by the following
property, which generalizes the invariants used in examples such as
currency transfers and resource tracking.

\begin{property}[Divisible Asset Conservation]
  \label{prop:divisible-asset-conservation}
  For each divisible asset $\stipula{a} \in \mathcal{A}$, declared
  with a fixed total quantity $\kappa_{\stipula{a}}$, the system
  ensures that the total amount held by all participants in
  $\mathcal{P}$ and the contract \stipula{C} remains constant:
  \[
    \sum_{X \in \mathcal{P} \cup \{ \stipula{C} \}} X.\stipula{a} = \kappa_{\stipula{a}}
  \]
\end{property}

\begin{example}
  The \stipula{Deposit} contract provides an illustration of
  Property~\ref{prop:divisible-asset-conservation}.  In this contract,
  the farmer transfers an amount \jml{h} of the divisible asset
  \jml{flour} to the contract by calling the \jml{send(h)} function.
  Importantly, this transfer does not create assets: the amount
  \jml{h} is subtracted from the farmer's \jml{Farm.flour} field and
  added to the contract's \jml{Deposit.flour} field.  This is
  reflected in the generated JML postcondition:
  
  {\small
\begin{lstlisting}[frame=none,numbers=none,language={[JML]Java}]
 /*@ public normal_behavior
    ...
    @ ensures     Deposit.flour == \old(Deposit.flour) + h 
    @             && Farm.flour == \old(Farm.flour) - h 
    @             && Client.flour == \old(Client.flour);
    ...
    @*/
    public final static void send(int h) {
        Deposit.flour = Deposit.flour + h;
        Farm.flour = Farm.flour - h;
    }
\end{lstlisting}
  } 
  
\noindent
Thus, the total quantity of the asset \jml{flour} remains constant
across the system, as required by the conservation property.  The
complete translation of the \stipula{Deposit} contract, including this
method, is shown in\iftoggle{extended}{
  Appendix~\ref{appendix:deposit}}{~\cite{extended-paper}}.
\end{example}

\paragraph{Functions.}
In {\Stipula}, contract behavior is defined by transitions that
correspond to functions and events.  Each transition is enabled under
specific conditions, such as the current control state, the invoking
party, and the availability of required assets. And it may update
fields, transfer assets, or trigger further state changes.  To reflect
this logic in Java, each {\Stipula} function is translated into a
static Java method with a formal specification in terms of a JML
contract:
\begin{itemize}
\item the \jml{requires} clause specifies the initial control state,
  party permissions, and preconditions over asset ownership and
  fields;
\item the \jml{ensures} clause describes the resulting state
  transition, updates to fields, and asset transfers;
\item the \jml{assignable} clause enumerates the state variables that
  may be modified.
\end{itemize}

Preconditions in the generated JML specifications are derived directly
from {\Stipula}'s guards, permissions, and asset requirements.  In
particular, they capture the enabling conditions for invoking a
function, rather than being reconstructed backwards from the
post-state.
\begin{example}
  \label{ex:verify-single}
  We illustrate the encoding of functions and states in JML with the
  \stipula{buy()} function in
  Listing~\ref{code:license-stipula}. Observe that the invariant for
  asset exclusivity is implicitly present in the pre- and
  postcondition. Fully automatic verification of the contract in
  {\KeY}, including invariant preservation, takes fractions of a
  second.
{\small
\begin{lstlisting}[frame=none,numbers=none,language={[JML]Java}]
/*@ public normal_behavior
  @ requires   License.balance && License.token;
  @ ensures    Licensor.balance && !License.balance &&
  @            Licensee.token && !License.token;
  @ assignable Licensor.balance, License.balance, Licensee.token, License.token;
  @*/
  public final static void buy() {
     Licensor.balance = true;
     License.balance = false;
     Licensee.token = true;
     License.token = false;
  }
\end{lstlisting}
}
\end{example}

\begin{definition}[Indivisible Asset Transfer Semantics]
  Let $\stipula{a} \in \mathcal{A}$ be an indivisible asset of
  contract \stipula{C}, and
  $\stipula{P}, \stipula{Q} \in \mathcal{P}$. A transfer of
  \stipula{a} from \stipula{P} to \stipula{Q} is modeled by the
  following postcondition:
  \[
    \lnot \stipula{P.a} \land \stipula{Q.a} \land \bigwedge_{\stipula{R} \in \mathcal{P} \cup
      \{\stipula{C}\} \setminus \{\stipula{P}, \stipula{Q} \}} (\stipula{R}.\stipula{a} = 
    \old{\stipula{R}.\stipula{a}}) 
  \]
  where $\old{\stipula{R.a}}$ refers to the value of $R.\stipula{a}$ 
  before the transfer is executed.
\end{definition}

This pattern guarantees that only the ownership of \stipula{P} and
\stipula{Q} changes, preserving the asset exclusivity invariant of
{\Stipula}. In addition, all changes are confined to the locations
occurring in the \jml{assignable} clause. If we assume
Property~\ref{prop:asset-exclusivity} to be ensured by the exclusivity
invariant, then \stipula{Q.a} alone is sufficient as postcondition.
Conditional statements and field updates in a {\Stipula} function
body are translated directly into Java code.

\subsection{Time and Events}
\label{sec:time-events}

Timed clauses in {\Stipula} define transitions that become enabled
only once logical time reaches a specified value. Such a transition is
written as
\[
  \stipula{now} + \stipula{k} \event \Qwithat \, \{ \, \stipula{S} \, \} \,\tostate\, \Qwithat'
\]
which expresses that, when time reaches $\stipula{now} + \stipula{k}$
and the current state is $\Q$, the contract may execute \stipula{S}
and move to state $\Q'$.  In the Java translation, these
time-dependent transitions are not realized by manipulating a global
clock. Instead, each timed event is guarded by a symbolic boolean
variable representing the time constraint. For example, a transition
scheduled at a future time is translated into code of the form:

\begin{lstlisting}[language={[JML]Java},frame=none,numbers=none]
if (ev_event1) { event1(); return; }
\end{lstlisting}

\noindent
Here, \texttt{ev\_event1} is a symbolic guard introduced by the
translator. Verification tools such as {\KeY} handle this guard
symbolically, thereby exploring both possibilities: one where the
event occurs and one where it does not. Each event is mapped to a
dedicated Java method, named systematically (\texttt{event1()},
\texttt{event2()}, $\ldots$), that encapsulates the corresponding body
$\stipula{S}$.\footnote{Deterministic event names are assigned via an
  internal counter to guarantee consistency and traceability.}  A key
point is that the symbolic guard \texttt{ev\_event1} does not encode
the source state $\Q$ explicitly. Instead, state constraints are
enforced structurally: our translator automatically generates
so-called \emph{scenario methods} (see Section~\ref{sec:scenarios}),
which represent feasible execution paths. An event method is reachable
only from the correct source state along such a path. In other words,
the interplay between scenario construction and method preconditions
guarantees that events can fire only in the intended states.

All event methods are statically defined during translation and never
generated dynamically at runtime.  This strategy eliminates the need
for explicit clocks or schedulers: time-dependent behavior is captured
entirely through symbolic guards. As a result, temporal reasoning can
be carried out using standard symbolic verification
tools. Furthermore, static analyses can identify and prune unreachable
event branches, improving efficiency. For instance, the technique
described in~\cite{Laneve2024} can be integrated into our translator,
and we plan to pursue this as future work.

\subsection{Cyclic Behavior and Loop Translation}
\label{sec:cyclic-behavior}
%
%
%
When translating a {\Stipula} contract whose underlying automaton
exhibits disjoint cycles, each cycle is mapped to a dedicated Java
\jml{while} loop. Conceptually, the body of the loop corresponds to a
single traversal of the cycle, \emph{i.e.}, one complete execution of
the contract operations contained within it. During each iteration,
the loop updates both the contract's asset variables and its
control-state fields to reflect the effect of the executed operations.
Loop execution is not governed by explicit counters hard-coded in the
program, but by symbolic scalar variables introduced during
translation. These variables represent iteration bounds and loop
counters, and they serve two complementary purposes:
\begin{enumerate}
\item \emph{operational control}: they determine when and how many
  times the loop body may be executed;
\item \emph{specification support}: they provide the basis for precise
  JML annotations, such as loop invariants and postconditions, that
  capture the intended effect of repeated executions.
\end{enumerate}

This combination ensures that the generated Java code remains faithful
to the original contract semantics while enabling deductive
verification tools to reason soundly about all possible iterations. By
structuring cycles as loops annotated with symbolic constraints, the
translation bridges the gap between the automaton view of contracts
and the logic-based reasoning frameworks used in verification.

%
%

\begin{example}
  \label{ex:JMLdeposit}
  Consider the \stipula{Deposit} contract in
  Listing~\ref{code:deposit-stipula}.  The translator generates scalar
  parameters \jml{h}, \jml{w}, \jml{h_send}, and \jml{counter} to
  represent single asset transfer amounts and the number of
  iterations, respectively.
  The generated code
  implements the cyclic behavior as follows: For instance, one part of
  the invariant states that the amount of \jml{flour} held by the
  contract evolves according to the sum of incoming deposits minus
  outgoing uses: {\small
\begin{lstlisting}[language={[JML]Java},frame=none,numbers=none]
  /*@ loop_invariant
    @ ... flour == \old(flour) - i * w/cost_flour + i * h_send; ...
    @*/
  while (i < counter) {
    buy(w);
    send(h_send);
    i++;
  }

\end{lstlisting}
  }
  
  Other parts of the invariant reflect the evolution of the farm and
  client asset fields (\jml{Farm.flour}, \jml{Client.flour}, etc.)
  and include standard loop annotations such as bounds and decreases
  clauses.  The translator automatically generates the full invariant
  by analyzing how each variable changes during loop execution (see\iftoggle{extended}{ Appendix~\ref{appendix:deposit} 
}{~\cite{extended-paper}} for the complete form).
\end{example}

It is important to emphasize a limitation of our current encoding
strategy.  Parameters such as \jml{w} in Example~\ref{ex:JMLdeposit}
are treated as symbolic constants throughout the entire execution of a
loop. In other words, all iterations are analyzed under a single
symbolic instantiation of the parameter. While the semantics of
{\Stipula} would, in principle, permit different values of \jml{w}
across different iterations, verifying such non-deterministic behavior
would require loop invariants that quantify over sequences of
iteration-dependent values. At present, verification tools such as
{\KeY} cannot discharge such invariants automatically. Supporting this
more general setting remains theoretically possible by allowing
user-supplied invariants and resorting to interactive proofs, but this
goes beyond the scope of our fully automated approach.

We also note that the loop in Example~\ref{ex:JMLdeposit} is bounded
by the symbolic parameter \jml{counter}, which provides an upper bound
on the number of iterations. Our construction associates a decreasing
variant, namely \jml{(counter - i)}, which enables {\KeY} to
automatically prove termination for contracts whose automata consist
of disjoint cycles. For more intricate cyclic structures, where such
simple variants are not available, proving termination would require
richer annotations and remains an open direction for future work.

Finally, regarding correctness, the loop invariants generated by our
translation suffice to establish preservation of the basic asset and
state properties in the examples we studied. In general, however,
automatically synthesized invariants are intentionally conservative:
they guarantee soundness but may be incomplete, in the sense that they
do not allow {\KeY} to prove every conceivable postcondition.  Our
design philosophy prioritizes invariants that are simple, structurally
derived, and always generated automatically. This choice ensures
robustness and automation for typical {\Stipula} contracts, while
leaving open the possibility of user-supplied annotations in cases
where more complex behaviors demand stronger reasoning power.

\subsection{Scenario-Based Reasoning}
\label{sec:scenarios}

To verify behavioral properties across complete execution paths within
a {\Stipula} contract, we declare \emph{scenario methods} that
represent legal sequences of contract actions. These methods model
complete paths through the underlying automaton of a {\Stipula}
contract, from initial to final states. Each such method is annotated
with a JML contract summarizing its overall effect, including field
updates, asset transfers, and control state progression.

\begin{example}\label{ex:scenario}
  The following scenario method models the successful completion of
  the contract in Listing~\ref{code:license-stipula}. It ensures that
  the assets are swapped as expected.  During verification, the
  already proven contracts of the called methods are used, the method
  bodies need not be inlined.

  {\small
\begin{lstlisting}[language={[JML]Java},frame=none,numbers=none]
  /*@ requires Licensor.token && Licensee.balance ; 
    @ ensures  Licensee.token && Licensor.balance ;*/
    public static void success() {
      offer(); activate(); buy();
    }
\end{lstlisting}}
\end{example}
%
%
%
%
%
%

When a {\Stipula} contract features disjoint cycles or branching
behavior that may lead to structurally distinct executions, our
translator automatically generates a dedicated scenario method for
each feasible path. Each scenario method corresponds to a linearized
execution trace, thereby capturing one possible evolution of the
{\Stipula} contract.  The generation process begins with a static
analysis of the contract's control structure.  In particular, the
translator verifies that the clauses of the contract give rise to
disjoint control paths. This is achieved by constructing the set of
linear traces induced by the underlying automaton and checking that
the cycles explored along each trace do not overlap.  Once this
structural property has been established, the translator can safely
emit one scenario per disjoint trace, with each scenario representing
a distinct and non-interfering contract behavior.  This design makes 
contract executions explicit for automated reasoning, 
and isolates traces into independent scenarios to avoid combinatorial explosion.

\begin{example}
  Consider a \stipula{Loan} contract that regulates a sequence of
  financial operations between a \stipula{Client} and a
  \stipula{Bank}.

  {\small
\begin{lstlisting}[language=stipula,frame=none,numbers=none,mathescape=true]
  @Start Bank : give_money()[w](w == amount) {
      w $\lolli$ Client ;
      now + 30 >> @Pay1 {  } $\tostate$ @Fail
  } $\tostate$ @Pay1

  @Start Bank : withdraw()[u](u == amount * interest_rate) {
      u $\lolli$ Client
      "The_Bank_withdraws" $\send$ Client;
  } $\tostate$ @Withdraw
\end{lstlisting}
}

In particular, the \stipula{Client} and the \stipula{Bank} agree on
the \stipula{amount} of the loan and the
\stipula{interest\_rate}. They also agree on a number of installment
payments and on a conditional early withdrawal by the
\stipula{Bank}. If the \stipula{Bank} exercises the early withdraw
then it has to pay a penalty to the client that is equal to
\stipula{amount * interest\_rate}.  (The complete code with three
payment installments is available in\iftoggle{extended}{
  Appendix~\ref{appendix:loan}}{~\cite{extended-paper}}.)  This
contract exhibits two disjoint execution paths: one where the
\stipula{Client} proceeds through all the payment steps, and another
where the \stipula{Bank} initiates an early withdrawal.  The two
execution paths are represented by the following two scenarios
automatically generated by the translator (variables
  \stipula{u}, \stipula{h}, \stipula{w} are bound in the precondition
  of the JML contract of the scenario, not shown here):

{\small
\begin{lstlisting}[language={[JML]Java},frame=none,numbers=none]
public final static void seq1() {           public final static void seq2() {
  give_money(w);                              withdraw(u);
  if (ev_event1) { event1(); return; }      }
  pay_installment1(h);
  if (ev_event2) { event2(); return; }
  pay_installment2(h);
  if (ev_event3) { event3(); return; }
  pay_installment3(h); }
\end{lstlisting}}
\end{example}

\section{Implementation and Evaluation}
\label{sect:casestudies}

To evaluate our translation and verification method, we implemented a
translator available in the online
repository~\cite{keystipulaprototype}.  The tool, realized in
ca.\ 2,000 lines of Java, uses ANTLR4 for parsing and a
listener-based traversal to extract contract components such as
parties, assets, fields, states, and transitions.  We assess its
effectiveness in this section by analyzing four representative
{\Stipula} contracts:
\begin{description}
\item[\textsf{Betting}:] illustrating branching resolution logic based
  on external outcomes;
\item[\textsf{Deposit}:] modeling recurring resource exchanges between
  client and provider, with timed fallback events and cyclic asset
  flows;
\item[\textsf{Loan}:] encoding installment-based repayment with symbolic
  arithmetic over loan parameters and time-triggered enforcement
  mechanisms;
\item[\textsf{License}:] involving timed obligations and conditional
  asset transfers.
\end{description}

These case studies encompass the main features of {\Stipula}
contracts, including asset transfer, exclusive ownership, timed
events, and cyclic interaction.  While not fully exhaustive, this
selection demonstrates that our approach handles the major constructs
found in practice.
Each {\Stipula} contract was automatically translated using our tool
chain (see~\iftoggle{extended}{Appendix~\ref{appendix:code}
}{\cite{extended-paper}} for the {\Stipula} source and generated
Java+JML code), and verified using {\KeY} in automatic mode.  No
manual proof steps or user-supplied annotations were required.  The
reported performance results were obtained on a MacBook Pro (2023)
equipped with an Apple M2 Pro processor and 16 GB of RAM.

We focus on generic contract properties that are automatically
generated together with the Java+JML code and verified without user
interaction.  These include functional correctness, loop invariants,
total termination, and symbolic handling of time-triggered
transitions.  Custom or domain-specific properties could be added
manually as additional postconditions, but are not required for the
scenarios we test.  The following verification goals are considered:

\begin{description}
\item[P1: Functional Correctness] Each scenario method satisfies the
  expected final state as specified by the automatically generated
  \jml{ensures} clauses.
\item[P2: Loop Termination] All loops are annotated with decreasing
  variant terms that are sufficient to prove termination.
\item[P3: Loop Invariant Preservation] Automatically generated
  inductive invariants ensure that key state relationships are
  preserved across iterations.
\item[P4: Time-Guard Soundness] Event-triggered clauses are guarded by
  symbolic boolean variables (e.g., \jml{ev_event1}), ensuring that
  transitions corresponding to timed clauses occur only under valid
  scheduling conditions.
\end{description}

{\small
\begin{table}
\centering
\begin{tabular}{l@{\hskip 12pt}c@{\hskip 12pt}c@{\hskip 12pt}c@{\hskip 12pt}c@{\hskip 12pt}|@{\hskip 12pt}c}
\toprule
\textbf{Contract} & \textbf{P1} & \textbf{P2} & \textbf{P3} & \textbf{P4} & \textbf{Time} \\
\midrule
\stipula{Betting} & \checkmark & -- & -- & \checkmark & $\sim$2.1s \\
\stipula{Deposit} & \checkmark & \checkmark & \checkmark & -- & $\sim$1.3s \\
\stipula{Loan} & \checkmark & -- & -- & \checkmark & $\sim$1.8s \\
\stipula{License} & \checkmark & -- & -- & \checkmark & $\sim$0.7s \\
  \bottomrule
  \\
\end{tabular}
\caption{Properties verified in the case studies. A dash ($"-"$) denotes that no property of the corresponding category applies to the given contract.}
\label{tab:verification-summary}
\end{table}
}

Table~\ref{tab:verification-summary} summarizes the successfully
verified and relevant properties for each contract.  Verification time
reflects the duration reported by {\KeY} when proving the main
top-level scenario. Verification times for other scenarios tend to be
somewhat less and the times to verify individual function and event
contracts are negligible.

For the \stipula{Betting} contract, the outcome logic is expressed
through mutually exclusive branches in its \jml{data()} method.
Verification ensures that all valid symbolic inputs result in safe
asset redistribution, and that execution halts in a consistent final
configuration.

In the \stipula{Deposit} contract, cyclic interactions between client
and provider are translated into a \jml{while} loop with a symbolic
iteration bound (\jml{counter}).  Loop invariants preserve the
consistency of transferred assets across rounds, while a variant term
guarantees termination.  Timed transitions are abstracted via symbolic
events (e.g., \jml{ev_event1}, \jml{ev_event2}), which may interrupt
execution early.

The \stipula{Loan} contract models installment-based repayment with
time constraints.  The translated code verifies that each installment
is transferred correctly and only when the corresponding symbolic time
guard holds.  The contract's structure relates to properties
\textbf{P1} and \textbf{P4}.

In the \stipula{License} contract, verification ensures that all
conditional obligations and asset transfers, such as license
activation or revocation, occur only under valid scheduling
conditions.  All possible contract outcomes are covered, based on
symbolic inputs like price, deadlines, and initial ownership.

\section{Related Work}
\label{sect:relatedworks}

The formal modeling and verification of digital contracts receives
increasing attention, especially at the intersection of legal
informatics, programming languages, and formal methods.  Our work
contributes by bridging a legal domain-specific language ({\Stipula})
with a deductive verification framework ({\KeY}) through translation
into JML-annotated Java.

Legal modeling frameworks like Ergo~\cite{ergo},
OpenLaw~\cite{OpenLaw}, Lexon~\cite{Lexon}, and Accord~\cite{Accord}
embed contracts into broader systems, but they lack a precise formal
semantics.  {\Stipula} offers an operational model with explicit
permissions, assets, and timed clauses.  While Catala~\cite{Catala}
formalizes legislative logic and has a runtime environment, it does
not address contract verification.

Closer to our work, prior efforts explored translation-based
verification: OCL-to-Java with JML~\cite{hamie2002ocl}, and
Circus-to-Java for formal reasoning~\cite{DBLP:conf/fm/FreitasC06}.
Our approach applies this paradigm to legal contracts, preserving
their normative and temporal semantics.

While the current {\Stipula} runtime is realized by
  compilation to Java~\cite{stipulaprototype}, the language is
implementation-agnostic and could be compiled to blockchain smart
contracts like Solidity~\cite{Solidity} or Obsidian~\cite{Obsidian}.
This path is promising, as our verification approach using
{\KeY}~\cite{DBLP:series/lncs/10001} can build on established work
that already applies the system to blockchain platforms like
Hyperledger Fabric~\cite{BeckertHerdaKirsten2018_1000092715} and
Solidity via the {\SolidiKeY} tool~\cite{DBLP:conf/isola/AhrendtB20}.
Using these advances would enable rigorous verification of {\Stipula}
contracts on decentralized infrastructures.


\section{Conclusion}
\label{sect:conclusion}

We presented a translation-based approach for verifying {\Stipula}
contracts by translating them into JML-annotated Java and applying
deductive verification.  This enables reasoning over normative
properties --- permissions, asset transfers, state transitions, and
timed clauses --- using an existing deductive verification tool.
What we verify in this setting is the correctness of the Java/JML
encoding of {\Stipula}'s semantics, derived automatically from
generation of scenarios, rather than an independent abstract notion of
functional correctness of contracts. Allowing users to state
additional derived properties directly in {\Stipula} would be a
natural extension, but it is beyond the scope of this work.
Alternative paradigms based on state-based formalisms (e.g.,
statecharts or timed automata) could in principle support reasoning in
temporal logics over possible behaviors. We chose a deductive setting
with {\KeY} to leverage its mature automation and direct JML support,
while exploring complementary verification approaches remains a
promising avenue for the future.

Our translation targets an expressive, yet analyzable, fragment of
{\Stipula}, with symbolic time, disjoint loops, loop-free clauses, and
limited non-determinism.  Within this fragment, verification is fully
automatic and requires no manual annotation or interactive proof.
Case studies confirm that functional and temporal behavior can be
verified compositionally, suggesting that legal contracts can be
verifiable by design if execution semantics is preserved.

The current approach synthesizes scenario methods by statically
traversing the contract automaton.  This relies on a structural
restriction: all cycles must be disjoint. Future work will focus on lifting this
restriction to support overlapping cycles and more general control
flow.  One possible direction is to introduce dynamic scheduling
mechanisms, such as a \emph{dispatch} table, that can track and
trigger enabled clauses at runtime.  While this would increase
expressiveness, it also introduces verification challenges that may
require interactive proofs or hybrid verification strategies.

Deductive tools like {\KeY} can often produce counterexamples, which,
when mapped back to the legal domain, help authors detect
inconsistencies and refine their contracts.  Another direction is
integrating runtime or hybrid verification to support cases where full
deductive reasoning is infeasible.

\subsubsection*{Acknowledgments}
We thank Maximilian Scheid for his work on the implementation of the
translator.

\bibliographystyle{splncs04}
\bibliography{references}
\iftoggle{extended}{
\newpage
\appendix

\section{Verifying  a Contract having Disjoint Cycles}
\label{appendix:cycles}

To verify that a contract has disjoint cycles, we construct all the
cycles and verify the absence of overlapping. To define the algorithm
we use the following auxiliary operations and notions:
\begin{itemize}
\item Clauses of a contract are identified by tuples
  $\clause{\Q}{\f}{\Q'}$ or $\clause{\Q}{\ev_n}{\Q'}$; we use $\H$ to
  range over $\f$ and $\ev_n$.  When $\C$ is a {\Stipula} contract, we
  write $\clause{\Q}{\H}{\Q'} \in \C$ when $\C$ has a clause
  $\clause{\Q}{\H}{\Q'}$;
\item \emph{linear traces} of clauses, ranged over by $\acomp$,
  $\acomp'$, $\ldots$, are sequences of clauses without repetitions
  $\clause{\Q_1}{\H_1}{\Q_2} \comp \ldots \comp
  \clause{\Q_{n-1}}{\H_n}{\Q_n}$;
\item
  $\states{\clause{\Q_1}{\H_1}{\Q_2} \comp \ldots \comp
    \clause{\Q_{n-1}}{\H_n}{\Q_n}} = \{ \Q_1, \cdots , \Q_n\}$;
\item the \emph{extension} $\acomp \addcomp \clause{\Q}{\H}{\Q'}$ of a
  linear trace $\acomp$ with the clause $\clause{\Q}{\H}{\Q'}$ is
  \[
    \begin{array}{rl}
      \acomp \addcomp \clause{\Q}{\H}{\Q'} \eqdef &
                                                    \left\{
                                                    \begin{array}{l@{\qquad\quad}l}
                                                      \acomp \comp \clause{\Q}{\H}{\Q'}
                                                      & \text{if } \acomp = \acomp' \comp \clause{\Q''}{\H'}{\Q} \; \text{ and } \;
                                                        \Q' \notin \states{\acomp}
                                                      \\
                                                      \acomp
                                                      & \text{otherwise}
                                                    \end{array} \right.
    \end{array}
  \]
  
  Observe that $\clause{\Q}{\H}{\Q'}$ is not added when either
  (\emph{i}) $\Q$ is not the last state of $\acomp$ or (\emph{ii})
  when $\acomp$ contains a cycle if $\clause{\Q}{\H}{\Q'}$ is added.
  We also write
  $\{ \acomp_1, \cdots , \acomp_n \} \addcomp \clause{\Q}{\H}{\Q'}
  \eqdef \{ \acomp_1\addcomp \clause{\Q}{\H}{\Q'}, \cdots , \acomp_n
  \addcomp \clause{\Q}{\H}{\Q'}\}$.
\end{itemize}

The algorithm uses two sets of linear traces
$\mathcal{A}, \mathcal{C}$ --- $\mathcal{A}$ contains linear traces
without cycles, $\mathcal{C}$ contains cyclic linear traces. Consider
a {\Stipula} contract $\C$ with initial state $\Q_0$.  We start with
$\mathcal{A}_0, \mathcal{C}_0$ where
\[
  \begin{array}{rl}
    \mathcal{A}_0 = & \bigl\{ \; \clause{\Q_0}{\H}{\Q} \quad | \quad \clause{\Q_0}{\H}{\Q} \in \C \; 
                      \texttt{ and } \; \Q_0 \neq \Q \; \bigr\}
    \\
    \mathcal{C}_0 = & \bigl\{ \; \clause{\Q_0}{\H}{\Q_0} \quad | \quad \clause{\Q_0}{\H}{\Q_0} \in \C  \; \bigr\}
  \end{array}
\]
and we iteratively define $\mathcal{A}_{i+1}, \mathcal{C}_{i+1}$ as follows
\[
  \begin{array}{rl}
    \mathcal{A}_{i+1} = & \bigl\{ \; \mathcal{A}_i \addcomp \clause{\Q}{\H}{\Q'} \quad | \quad \clause{\Q}{\H}{\Q'} \in \C  \; \bigr\}
    \\
    \mathcal{C}_{i+1} = & \mathcal{C}_i \bigcup \;
                          \bigl\{ \; \clause{\Q}{\H}{\Q} \quad | \quad \clause{\Q}{\H}{\Q} \in \C \; \text{
                          and } \; \acomp \comp \clause{\Q'}{\H'}{\Q} \in \mathcal{A}_i \; \bigr\}
    \\
                        & \quad \bigcup \;
                          \bigl\{ \; \acomp' \comp \clause{\Q}{\H}{\Q'} \quad | \quad \clause{\Q}{\H}{\Q'} \in \C \; 
                          \text{ and } \; \acomp \comp \acomp' \in \mathcal{A}_i
    \\
                        & \qquad \qquad \qquad \qquad \quad \text{ and } \; \acomp' = \clause{\Q'}{\H'}{\Q''} \comp \acomp'' \comp \clause{\Q'''}{\H''}{\Q}  \; \bigr\}
  \end{array}
\]

Observe that $\mathcal{C}_{i+1}$ extends $\mathcal{C}_{i}$ by adding
cycles of length one on final states of linear traces of
$\mathcal{A}_{i}$ and longer cycles if
$\acomp \comp \acomp' \in \mathcal{A}_i$ and
$\acomp' \comp \clause{\Q}{\H}{\Q'}$ is a cycle.

The above iteration always terminates --- there exists $n$ such that
$\mathcal{A}_{n}, \mathcal{C}_{n} = \mathcal{A}_{n+1},
\mathcal{C}_{n+1}$, because the clauses of a contract are finite and a
contract may have finitely many linear traces.  Upon termination, we
verify whether $\states{\acomp} \cap \states{\acomp'} = \varnothing$,
for all pairwise different linear traces
$\acomp, \acomp' \in \mathcal{C}_n$. If this is the case the contract
has disjoint states and we may verify it in automatic way.

\clearpage

\section{Case Study Code Listings}
\label{appendix:code}

\subsection{License Contract}
\label{appendix:license}
The \stipula{License} contract models a software licensing agreement with a time-limited offer and a trial period. 
The agreement involves two parties: a \stipula{Licensor} who owns the software license, 
and a \stipula{Licensee} who wishes to purchase it at a specific \stipula{cost}.
The contract's logic progresses through distinct states, 
from the initial offer to the final purchase or expiration. 
The process begins in the \stipula{@Inactive} state and proceeds as follows:
\begin{itemize}
\item The \stipula{Licensor} initiates the process by calling the \stipula{offerLicense()} function. This action deposits an indivisible license \stipula{token} into the contract's escrow and sets an activation \stipula{code}. A timer, \stipula{t_start}, begins, giving the \stipula{Licensee} a limited time to accept the offer. If the offer expires, the \stipula{token} is returned to the \stipula{Licensor}, and the contract terminates. Upon a successful offer, the state transitions to \stipula{@Proposal}.
\item While in the \stipula{@Proposal} state, the \stipula{Licensee} can call \stipula{activateLicense()} by paying the agreed-upon \stipula{cost} into escrow. In return, the \stipula{Licensee} receives the activation \stipula{code}. This starts a new timer, \stipula{t_limit}, which defines the trial period, and moves the state to \stipula{@Trial}. If this trial period expires, the \stipula{Licensee} is automatically refunded their payment, the \stipula{token} is returned to the \stipula{Licensor}, and the contract ends.
\item During the \stipula{@Trial} period, the \stipula{Licensee} can finalize the purchase by calling the \stipula{buy()} function. This action executes the final transfer: the escrowed payment (\stipula{balance}) is sent to the \stipula{Licensor}, and the escrowed license \stipula{token} is sent to the \stipula{Licensee}, successfully completing the transaction and moving the contract to the \stipula{@End} state.
\end{itemize}
{\small 
\begin{lstlisting}[language=stipula,caption={Original \smallang\ code
  for the \stipula{License}
  contract},label={lst:license-stipula},frame=none,numbers=none]
stipula License {
    assets balance, token
    fields t_start, t_limit, cost, code
    agreement (Licensor, Licensee)(t_start, t_limit, cost){
		Licensor , Licensee : t_start, t_limit, cost
    } => @Inactive
    @Inactive Licensor : offerLicense (x)[n]{
        n -o token
        x -> code
        now + t_start >> @Proposal{
                token -o Licensor
            } => @End
    } => @Proposal
    @Proposal Licensee : activateLicense()[b] (b == cost) {
        b -o balance
        code -> Licensee
        now + t_limit >> @Trial {
                balance -o Licensee
                token -o Licensor
            } => @End
    } => @Trial
    @Trial Licensee : buy ()[]{
        balance -o Licensor
        token -o Licensee
    } => @End
}
\end{lstlisting}}
\newpage
{\small 
\begin{lstlisting}[language={[JML]Java}, caption={Generated Java code
  for the \stipula{License} contract},
label={lst:appendix-license},frame=none,numbers=none]
public final class License {

    public static int Licensor, Licensee;
    public static boolean LicensorBalance, LicenseeBalance;
    public static boolean balance, token;
    public static boolean LicensorToken, LicenseeToken;
    public static int t_start, t_limit;
    public static int cost;
    public static int code, LicenseeCode;
    public static int b, x, n;
    public static boolean ev_event1, ev_event2;

    /*@ public normal_behavior
      @ ensures     LicensorToken == false && token == true && code == x;
      @ assignable  LicensorToken, code, token;
      @*/
    public final static void offerLicense(int x, boolean n) {
        token = true;
        LicensorToken = false;
        code = x;
    }

    /*@ public normal_behavior
      @ requires    b==cost;
      @ ensures     balance == true && LicenseeBalance == false && LicenseeCode == code;
      @ assignable  LicenseeBalance, LicenseeCode, balance;
      @*/
    public final static void activateLicense(boolean b) {
        balance = true;
        LicenseeBalance = false;
        LicenseeCode = code;
    }

    /*@ public normal_behavior
      @ ensures     balance == false && LicensorBalance == true && LicenseeToken == true && token == false;
      @ assignable  LicenseeToken, LicensorBalance, balance, token;
      @*/
    public final static void buy() {
        LicensorBalance = true;
        balance = false;
        LicenseeToken = true;
        token = false;
    }

    /*@ public normal_behavior
      @ ensures     LicensorToken == true && token == false;
      @ assignable LicensorToken, token;
      @*/
    public final static void event1() {
        LicensorToken = true;
        token = false;
    }

    /*@ public normal_behavior
      @ ensures     balance == false && LicenseeBalance == true && LicensorToken == true && token == false;
      @ assignable LicenseeBalance, LicensorToken, balance, token;
      @*/
    public final static void event2() {
        LicenseeBalance = true;
        balance = false;
        LicensorToken = true;
        token = false;
    }

    /*@ public normal_behavior
      @ requires b==cost;
      @ ensures (
      @   (ev_event1 && LicensorToken == false && token == true && code == x)
      @   || (!ev_event1 && ev_event2 && LicensorToken == false && token == true && code == x && balance == true && LicenseeBalance == false && LicenseeCode == code)
      @   || (!ev_event1 && !ev_event2 && LicensorToken == false && token == true && code == x && balance == true && LicenseeBalance == false && LicenseeCode == code && LicensorBalance == true && LicenseeToken == true)
      @ );
      @*/
    public final static void seq1() {
        offerLicense(x, n);
        if(ev_event1){
            event1();
            return;
        }
        activateLicense(b);
        if(ev_event2){
            event2();
            return;
        }
        buy();
    }

}
\end{lstlisting}}
\newpage

\subsection{Betting Contract}
\label{appendix:betting}
The \stipula{Betting} contract, displayed in Listing~\ref{lst:betting-stipula}, 
provides an example of conditional asset transfers. 
In simple terms, this contract functions as a \emph{trusted, automated escrow service} for a bet between two parties, 
\stipula{Better1} and \stipula{Better2}, 
with a \stipula{DataProvider} acting as the \emph{oracle} to determine the outcome.

In the contract, each participant sequentially calls the \texttt{place\_bet} function. 
This function serves two crucial roles: 
first, it registers the participant's prediction (e.g., who will win a match); 
second, it transfers the participant's stake into the secure escrow assets controlled by the resolution phase, 
the \stipula{DataProvider} submits the actual outcome by calling the \stipula{data} function. 
The contract then executes its core logic: an \stipula{if-else} block that compares the official outcome to the predictions. 
Based on the results, the funds held in \stipula{wallet1} and \stipula{wallet2} are automatically redistributed to the winner(s) or, if neither party guessed correctly, to the \stipula{DataProvider}.

{\small 
\begin{lstlisting}[language=stipula,caption={Original \smallang\ code
  for the \stipula{Betting}
  contract},label={lst:betting-stipula},frame=none,numbers=none]
stipula Bet {
	assets wallet1, wallet2
	fields val1, val2, event, amount, t_before, t_after

	agreement (Better1 , Better2 , DataProvider)
			  (event, amount, t_before, t_after){ 
		Better1, Better2 : event, amount, t_before, t_after
	} => @Init

	@Init Better1: place_bet(x)[h] (h == amount){
		h -o wallet1
		x -> val1
		t_before >> @First { wallet1 -o Better1 } => @Fail
	} => @First

	@First Better2: place_bet(x)[h] (h == amount){
		h -o wallet2
		x -> val2
		t_after >> @Run {
			wallet1 -o Better1
			wallet2 -o Better2
		} => @Fail
	} => @Run

	@Run DataProvider: data(x,z)[](x==event){
		if (z==val1 && z==val2){
			wallet1 -o Better1
			wallet2 -o Better2
		}
		else if (z==val1 && z!=val2){
			wallet2 -o Better1
			wallet1 -o Better1
		}
		else if (z!=val1 && z==val2){
			wallet1 -o Better2
			wallet2 -o Better2
		}
		else{
			wallet2 -o DataProvider
			wallet1 -o DataProvider
		}
	} => @End
}
\end{lstlisting}}

{\small 
\begin{lstlisting}[language={[JML]Java}, caption={Generated Java code
  for the \stipula{Betting} contract},
label={lst:appendix-bet},frame=none,numbers=none]
public final class Bet {

    public static int Better1, Better2, DataProvider;
    public static boolean wallet1, wallet2;
    public static boolean Better1Wallet1, Better2Wallet1, DataProviderWallet1;
    public static boolean Better1Wallet2, Better2Wallet2, DataProviderWallet2;
    public static int val1, val2, event, amount;
    public static int t_before, t_after;
    public static int x, h, z;
    public static boolean ev_event1, ev_event2;

    /*@ public normal_behavior
      @   requires    h == amount;
      @   ensures     Better1Wallet1 == false && wallet1 == true && val1 == x;
      @   assignable  Better1Wallet1, val1, wallet1;
      @*/
    public final static void place_bet(int x, boolean h) {
        wallet1 = true;
        Better1Wallet1 = false;
        val1 = x;
    }

    /*@ public normal_behavior
      @   requires    h == amount;
      @   ensures     Better2Wallet2 == false && wallet2 == true && val2 == x;
      @   assignable  Better2Wallet2, val2, wallet2;
      @*/
    public final static void place_bet1(int x, boolean h) {
        wallet2 = true;
        Better2Wallet2 = false;
        val2 = x;
    }

    /*@ public normal_behavior
      @   requires    x == event;
      @   ensures
      @     ( (z == val1 && z == val2) &&
      @       (Better1Wallet1 && !wallet1 && Better2Wallet2 && !wallet2) )
      @     ||
      @     ( (z == val1 && z != val2) &&
      @       (Better1Wallet2 && !wallet2 && Better1Wallet1 && !wallet1) )
      @     ||
      @     ( (z != val1 && z == val2) &&
      @       (Better2Wallet1 && !wallet1 && Better2Wallet2 && !wallet2) )
      @     ||
      @     ( (z != val1 && z != val2) &&
      @       (DataProviderWallet2 && !wallet2 && DataProviderWallet1 && !wallet1) );
      @   assignable
      @     Better1Wallet1, Better1Wallet2, Better2Wallet1, Better2Wallet2,
      @     DataProviderWallet1, DataProviderWallet2, wallet1, wallet2;
      @*/
    public final static void data(int x, int z) {
        if (z == val1 && z == val2) {
            Better1Wallet1 = true;
            wallet1 = false;
            Better2Wallet2 = true;
            wallet2 = false;
        } else if (z == val1 && z != val2) {
            Better1Wallet2 = true;
            wallet2 = false;
            Better1Wallet1 = true;
            wallet1 = false;
        } else if (z != val1 && z == val2) {
            Better2Wallet1 = true;
            wallet1 = false;
            Better2Wallet2 = true;
            wallet2 = false;
        } else {
            DataProviderWallet2 = true;
            wallet2 = false;
            DataProviderWallet1 = true;
            wallet1 = false;
        }
    }

    /*@ public normal_behavior
      @   assignable Better1Wallet1, wallet1;
      @*/
    public final static void event1() {
        Better1Wallet1 = true;
        wallet1 = false;
    }

    /*@ public normal_behavior
      @   assignable Better1Wallet1, Better2Wallet2, wallet1, wallet2;
      @*/
    public final static void event2() {
        Better1Wallet1 = true;
        wallet1 = false;
        Better2Wallet2 = true;
        wallet2 = false;
    }

    /*@ public normal_behavior
      @   requires h == amount &&
      @            x == event;
      @   ensures
      @     ( (z == val1 && z == val2) &&
      @       (Better1Wallet1 && !wallet1 && Better2Wallet2 && !wallet2) )
      @     ||
      @     ( (z == val1 && z != val2) &&
      @       (Better1Wallet2 && !wallet2 && Better1Wallet1 && !wallet1) )
      @     ||
      @     ( (z != val1 && z == val2) &&
      @       (Better2Wallet1 && !wallet1 && Better2Wallet2 && !wallet2) )
      @     ||
      @     ( (z != val1 && z != val2) &&
      @       (DataProviderWallet2 && !wallet2 && DataProviderWallet1 && !wallet1) );
      @*/
    public final static void seq1() {
        place_bet(x, h);
        if (ev_event1) {
            event1();
            return;
        }
        place_bet1(x, h);
        if (ev_event2) {
            event2();
            return;
        }
        data(x, z);
    }
}
\end{lstlisting}}

\newpage
\subsection{Deposit Contract}
\label{appendix:deposit}
The \stipula{Deposit} contract, displayed in Listing~\ref{lst:deposit-stipula}, 
models a recurring supply-and-purchase agreement, 
illustrating how Stipula handles cyclic behavior and divisible assets. 
One party, the \stipula{Farm}, can stock a resource (\stipula{flour}), 
and another party, the \stipula{Client}, can purchase it at a pre-agreed price.

The contract's logic is built around an alternating cycle between two main states, 
\stipula{@RunF} (the Farm's turn) and \stipula{@RunC} (the Client's turn). 
The \stipula{Farm} makes an initial deposit with the \stipula{begin()} function. 
Then, the cycle proceeds as follows:
\begin{itemize}
    \item When in state \stipula{@RunC}, the \stipula{Client} can call the \stipula{buy()} function to make a purchase. The contract automatically calculates the amount of \stipula{flour} to dispense based on the payment \stipula{w} and the agreed \stipula{cost\_flour}. It then transfers the \stipula{flour} to the \stipula{Client} and routes the payment to the \stipula{Farm}. After a purchase, the state transitions to \stipula{@RunF}.
    \item When in state \stipula{@RunF}, the \stipula{Farm} can replenish the stock by calling the \stipula{send()} function, depositing more \stipula{flour} into the contract's escrow. This transitions the state back to \stipula{@RunC}, allowing the \stipula{Client} to buy again.
\end{itemize}
This cycle of sending and buying can repeat indefinitely. 
To ensure termination, the contract includes a 365-day deadline. 
A timed event ensures that if the contract is still active after this period, 
any remaining \stipula{flour} is safely returned to the \stipula{Farm}.

{\small 
\begin{lstlisting}[language=stipula,caption={Original \smallang\ code
  for the \stipula{Deposit}
  contract},label={lst:deposit-stipula},frame=none,numbers=none]
stipula Deposit {
  assets flour
  fields cost_flour

  agreement (Client, Farm)(cost_flour) {
    Client, Farm : cost_flour
  } => @Start

  @Start Farm : begin()[h]{
    h -> Client
    h -o flour
    now + 365 >> @RunF { flour -o Farm } => @End
    now + 365 >> @RunC { flour -o Farm } => @End
  } => @RunC

  @RunF Farm : send()[h]{
    h -> Client
    h -o flour
  } => @RunC

  @RunC Client : buy()[w](w/cost_flour <= flour){
    (w/cost_flour) -o flour, Client
    w -o Farm
  } => @RunF
}
\end{lstlisting}}

{\small 
\begin{lstlisting}[language={[JML]Java}, caption={Generated Java code
  for the \stipula{Deposit} contract},
label={lst:appendix-deposit},frame=none,numbers=none]
public final class Deposit {

    public static int Client, Farm;
    public static int ClientFlour, FarmFlour;
    public static int flour;
    public static int cost_flour;
    public static int h;
    public static int c_w, c_h;
    public static int counter;
    public static boolean ev_event1, ev_event2;

    /*@ public normal_behavior
      @ ensures     flour == \old(flour)  + h && FarmFlour == \old(FarmFlour)  - h;
      @ assignable  FarmFlour, flour;
      @*/
    public final static void begin(boolean h) {
        flour += h;
        FarmFlour -= h;
    }

    /*@ public normal_behavior
      @ ensures     flour == \old(flour)  + h && FarmFlour == \old(FarmFlour)  - h;
      @ assignable  FarmFlour, flour;
      @*/
    public final static void send(boolean h) {
        flour += h;
        FarmFlour -= h;
    }

    /*@ public normal_behavior
      @ requires    w/cost_flour<=flour;
      @ ensures     ClientFlour == \old(ClientFlour)  + (w/cost_flour) && flour == \old(flour)  - (w/cost_flour) && Farm == \old(Farm)  + w && Client == \old(Client)  - w;
      @ assignable  Client, ClientFlour, Farm, flour;
      @*/
    public final static void buy(boolean w) {
        ClientFlour += (w*1 / cost_flour);
        flour -= (w*1 / cost_flour);
        Farm += w;
        Client -= w;
    }

    /*@ public normal_behavior
      @ ensures     flour == false && FarmFlour == true;
      @ assignable FarmFlour, flour;
      @*/
    public final static void event1() {
        FarmFlour += flour;
        flour = 0;
    }

    /*@ public normal_behavior
      @ ensures     flour == false && FarmFlour == true;
      @ assignable FarmFlour, flour;
      @*/
    public final static void event2() {
        FarmFlour += flour;
        flour = 0;
    }

    /*@ public normal_behavior
      @ requires counter >= 0;
      @ ensures Client == \old(Client)+ counter * (- w) &&
      @         ClientFlour == \old(ClientFlour)+ counter * (+ (w/cost_flour)) &&
      @         Farm == \old(Farm)+ counter * (+ w) &&
      @         FarmFlour == \old(FarmFlour)- h + counter * (- h) &&
      @         flour == \old(flour)+ h + counter * (- (w/cost_flour)  + h);
      @*/
    public final static void seq1() {
        begin(h);
        if(ev_event1){
            event1();
            return;
        }
        if(ev_event2){
            event2();
            return;
        }
        int i=0;
        /*@ loop_invariant 0 <= i && i <= counter;
          @ loop_invariant c_w/cost_flour<=flour;
          @ loop_invariant ClientFlour == \old(ClientFlour) + i * (+ (c_w/cost_flour));
          @ loop_invariant flour == \old(flour) + i * (- (c_w/cost_flour)  + c_h);
          @ loop_invariant Farm == \old(Farm) + i * (+ c_w);
          @ loop_invariant Client == \old(Client) + i * (- c_w);
          @ loop_invariant FarmFlour == \old(FarmFlour) + i * (- c_h);
          @ assignable Client, ClientFlour, Farm, FarmFlour, flour, i;
          @ decreases counter - i;
         @*/
        while(i < counter) {
            buy(c_w);
            send(c_h);
            i++;
        }
    }

}
\end{lstlisting}}
\newpage

\subsection{Loan Contract}
\label{appendix:loan}
The \stipula{Loan} contract models a fixed-term installment loan between two parties. 
A \stipula{Bank} provides an initial loan to a \stipula{Client}, 
who must then repay it in three installments based on a pre-agreed \stipula{amount} and \stipula{interest_rate}.

The contract's logic follows a primary success path from the initial state, 
\stipula{@Start}, to the final state, 
\stipula{@Success}, but also includes alternative paths for cancellation or failure. 
The process begins when the \stipula{Bank} initiates the loan:
\begin{itemize}
\item In the \stipula{@Start} state, the \stipula{Bank} calls the \stipula{give_money()} function, transferring the principal amount \stipula{w} to the \stipula{Client}. This action initiates a 30-day timer and transitions the contract to the first repayment state, \stipula{@Pay1}.
\item The \stipula{Client} is then expected to make three repayments by calling the \stipula{pay_installment()} function. Each call transfers an installment payment \stipula{h}, calculated from the principal and interest rate, back to the \stipula{Bank}. This process moves the contract sequentially through the states \stipula{@Pay1}, \stipula{@Pay2}, and \stipula{@Pay3}.
\item After the third and final installment, the contract successfully concludes by moving to the \stipula{@Success} state. Each repayment step is governed by a 30-day timeout; if the \stipula{Client} fails to make a payment within the allotted time, the contract transitions to a \stipula{@Fail} state.
\end{itemize}
As an alternative to initiating the loan, the \stipula{Bank} has the option to call \stipula{withdraw()} from the \stipula{@Start} state, 
effectively canceling the agreement and moving to the \stipula{@Withdraw} state.

{\small 
\begin{lstlisting}[language=stipula,caption={Original \smallang\ code
  for the \stipula{Loan}
  contract},label={lst:loan-stipula},frame=none,numbers=none]
stipula Loan {
    fields interest_rate, amount

    agreement (Client, Bank)(interest_rate, amount) {
    	Client, Bank: interest_rate, amount
    } => @Start

    @Start Bank: give_money()[w](w == amount){
    	w -o Client
    	now + 30 >> @Pay1 { } => @Fail
    } => @Pay1

    @Pay1 Client: pay_installment()[h](h == (amount * (1 + interest_rate))/3){
    	h -o Bank
    	now + 30 >> @Pay2 { } => @Fail
    } => @Pay2

    @Pay2 Client: pay_installment()[h](h == (amount * (1 + interest_rate))/3){
    	h -o Bank
    	now + 30 >> @Pay3 { } => @Fail
    } => @Pay3

    @Pay3 Client: pay_installment()[h](h == (amount * (1 + interest_rate))/3){
    	h -o Bank
    } => @Success

    @Start Bank: withdraw()[]{
    	"The Bank withdraws" -> Client
    } => @Withdraw
}
\end{lstlisting}}

{\small 
\begin{lstlisting}[language={[JML]Java}, caption={Generated Java code
  for the \stipula{Loan} contract},
label={lst:appendix-loan},frame=none,numbers=none]
public final class Loan {

    public static int Client, Bank;
    public static int interest_rate;
    public static int amount;
    public static int w, h;
    public static boolean ev_event1, ev_event2, ev_event3;

    /*@ public normal_behavior
      @ requires    w==amount;
      @ ensures     Bank == \old(Bank)  - w && Client == \old(Client)  + w;
      @ assignable  Bank, Client;
      @*/
    public final static void give_money(boolean w) {
        Client += w;
        Bank -= w;
    }

    /*@ public normal_behavior
      @ requires    h==(amount*(1+interest_rate))/3;
      @ ensures     Bank == \old(Bank)  + h && Client == \old(Client)  - h;
      @ assignable  Bank, Client;
      @*/
    public final static void pay_installment(boolean h) {
        Bank += h;
        Client -= h;
    }

    /*@ public normal_behavior
      @ requires    h==(amount*(1+interest_rate))/3;
      @ ensures     Bank == \old(Bank)  + h && Client == \old(Client)  - h;
      @ assignable  Bank, Client;
      @*/
    public final static void pay_installment1(boolean h) {
        Bank += h;
        Client -= h;
    }

    /*@ public normal_behavior
      @ requires    h==(amount*(1+interest_rate))/3;
      @ ensures     Bank == \old(Bank)  + h && Client == \old(Client)  - h;
      @ assignable  Bank, Client;
      @*/
    public final static void pay_installment2(boolean h) {
        Bank += h;
        Client -= h;
    }

    /*@ public normal_behavior
      @*/
    public final static void withdraw() {
    }

    /*@ public normal_behavior
      @*/
    public final static void event1() {
    }

    /*@ public normal_behavior
      @*/
    public final static void event2() {
    }

    /*@ public normal_behavior
      @*/
    public final static void event3() {
    }

    /*@ public normal_behavior
      @ requires w==amount &&
      @          h==(amount*(1+interest_rate))/3;
      @ ensures (
      @   (ev_event1 && Bank == \old(Bank)  - w && Client == \old(Client)  + w)
      @   || (!ev_event1 && ev_event2 && Bank == \old(Bank)  - w  + h && Client == \old(Client)  + w  - h)
      @   || (!ev_event1 && !ev_event2 && ev_event3 && Bank == \old(Bank)  - w  + h  + h && Client == \old(Client)  + w  - h  - h)
      @   || (!ev_event1 && !ev_event2 && !ev_event3 && Bank == \old(Bank)  - w  + h  + h  + h && Client == \old(Client)  + w  - h  - h  - h)
      @ );
      @*/
    public final static void seq1() {
        give_money(w);
        if(ev_event1){
            event1();
            return;
        }
        pay_installment(h);
        if(ev_event2){
            event2();
            return;
        }
        pay_installment1(h);
        if(ev_event3){
            event3();
            return;
        }
        pay_installment2(h);
    }

    /*@ public normal_behavior
      @*/
    public final static void seq2() {
        withdraw();
    }

}
\end{lstlisting}}
}{}
\end{document}
